\documentclass[]{article}

\def\ben{\begin{equation}}
\def\een{\end{equation}}
\def\bea{\begin{eqnarray}}
\def\eea{\end{eqnarray}}
\input amssym.def
\input amssym.tex
\begin{document}

\hfuzz=100pt
\title{The Maximum Tension Principle in General Relativity  }
\author{G W Gibbons
\\
D.A.M.T.P.,
\\ Cambridge University,
\\ Wilberforce Road,
\\ Cambridge CB3 0WA,
 \\ U.K.}
\maketitle

\begin{abstract}
I suggest that classical General Relativity in four spacetime
dimensions  incorporates a Principal of Maximal Tension and give
arguments to show that the value of the maximal tension is $c^4
\over 4 G$. The relation of this principle to other, possibly
deeper, maximal principles is discussed, in particular the
relation to the tension in string theory.
In that case it leads to a purely classical
 relation between $G$ and the classical
string coupling constant  $\alpha ^\prime$ and the velocity of light
$c$ which does not involve Planck's constant.

\end{abstract}

\vfill \eject

\section{Introduction}
Jacob Bekenstein has always been interested in simple Physical
Principles (capitals intentional) and so it seems appropriate to
celebrate the 30th anninversary of his work
on black hole thermodynamics with an account of a simple, but
perhaps un-noticed, principle in classical general relativity
which, like the Christadoulous's idea of irreducible mass
and Hawking's area increase theorem,
also seems to point to to deeper things. The principle is one of
those "impossibility statements" that impose an upper (or lower)
bound on some physical quantity. The most obvious example is the
upper bound on velocity in Special Relativity. Another example is
the lower bound on temperature first noticed in kinetic theory and
now accepted as universal and embedded into the more general
framework of Statistical Mechanics. In like fashion one can
contemplate, and I shall,  a more general framework including
General Relativity as a special limit in which the Maximum Tension
Principle is embedded in a fundamental way. A striking feature of
the Principle is that while it has an  analogue  in higher
dimensions it takes on its most simple and natural form  in four
spacetime dimensions. Specifically I propose:

\medskip \noindent {\bf The Principle of Maximum Tension} {\it  The tension or
force between two bodies cannot exceed

\ben
F_g = {c^4 \over 4G}.
\een

}
\medskip

The number 4 seems to be correct from the examples I am about to
give but it may be  subject to revision in the light of future
developments. Numerically
\ben
F_g \approx 3.25  \times 10 ^{43} \thinspace  {\rm Newtons} , \een
which is about $3 \times 10 ^{39}
\thinspace {\rm Tonnes}$.
In support of my contention that it has gone relatively  unknown
among the relativity community, it is interesting to note that
 $c^4 \over G$ does not appear
among the  ``useful combinations'' of the two fundamental
constants, the velocity of $c$ and Newton's constant of gravitation
$G$
in the well known textbook \cite{Gravitation}.

At the Newtonian level of course the tension in the gravitational
field is unbounded. In the language developed by Maxwell,
the Newtonian stress tensor has the opposite properties from
those in electrostatics. It is  given by
\ben
T_{ij}= -{1 \over 4 \pi G} \Bigl [ \partial _i U \partial _j U- {
1\over 2} \delta _{ij} |\partial U |^2 \Bigr ],
\een
and the gravitational force per unit volume  $F_i= \partial _j T_{ij}$.
The force per unit area along the field lines, assumed to point along
the 1 direction for example,  is {\sl repulsive}
\ben
T_{11}= -{ 1\over 8 \pi G} | \partial U |^2
\een
while in the transverse direction there are tensions, not pressures:
\ben
T_{22}=T_{33}= + {1 \over 8 \pi G} |\partial U |^2.
\een
Maxwell and those that followed him, for example \cite{Lodge},
found this paradoxical: the more so because
the forces and tensions  between heavenly bodies, which were at that
time thought to be exerted through the ether, are so large
compared with what is encountered in an ordinary terrestrial
medium. In General Relativity
these stresses remain just as large  (very large)
for ordinary celestial bodies but they cannot become unbounded.
There is a natural limit because of the phenomenon of gravitational
collapse
and black hole formation.

My initial qualitative argument is too crude to
deliver a precise upper bound, merely an order of magnitude,
 but more sophisticated calculations do.
Consider two  bodies (possibly black holes, but not necessarily so)
of positive masses $M_1$ and $M_2$ separated a distance $D$ apart,
According to Newtonian theory, The gravitational force between them
is
\ben
F= {G M_1 M_2 \over D^2}= \bigl ({ GM_1 \over c^2D} \bigr ) \bigl
( {GM_2 \over c^2 D} \bigr ) {c^4 \over G} .\een
However $M_1 M_2$ cannot exceed ${1  \over4 } (M_1 + M_2)^2 $ and
therefore
\ben
F \le \bigl ( {M_1 +M_2 \over c^2 D} \bigr ) ^2 {c^4 \over 4G}.
\een
Now if there are to be two bodies rather than a single black hole
hole it must be true that $M_1 +M_2 < c^2 D$ and so the Principle
holds in this case. Obviously I have been a little  cavalier with
factors in the last sentence and so it is good to have some  more
exact arguments. The point is that  to keep the bodies apart we
need to pull them away from each other with some sort of strings.
In the axisymmetric case at least, and in the thin string limit, these
may be approximated by  conical defects running off to infinity
along the two portions  of the axis on the outer side of each
body.  The deficit angle $\delta$ given by
\ben
\delta = {8 \pi G \over c^4} F,
\een
where the force $F$ may  identified (at least in four spacetime
dimensions) with the tension or energy per unit length. Because
the deficit angle $\delta$  cannot exceed $2 \pi$ we again see
that the Principle holds. Note that this upper bound for the tension
should be contrasted with a Bogomol'nyi style lower bound
established in \cite{GibbonsComtet}.

To see the bound operating  in more detail, particularly  in the case of black
holes, we consider static axi-symmetric vacuum metrics and use what
may be called

  \section{The Method of Newtonian Rods} This arose out a paper of
Einstein and Rosen
\cite{EinsteinRosen}
and was  applied to black holes in
\cite{IsraelKhan,Gibbons,Gibbons2}. The method  has been nicely reviewed
recently in \cite{Dowker} and so I will not give many mathematical
details here. We start by following Weyl and expressing  a static
axisymmetric vacuum  metric as
\ben
ds^2 = - e^{2U} dt^2 + e^{-2U} \Bigl \{  e^{2k} \bigl ( d z^2 + d
\rho^2 \bigl ) + \rho ^2 d \phi ^2 \Bigr \},
\een
where $U$ is an axisymmetric harmonic function on flat three
dimensional Euclidean space ${\Bbb E} ^3 $ with cylindrical
coordinates $(z, \rho, \phi)$ and the axisymmetric function $k$ is
given in terms of $ U$ by a line integral whose
contour-independence is a consequence of Laplace's equation for
$U$.

To obtain a single  black hole we choose for $U$ the Newtonian
potential of a uniform rod of mass $M$  length $L$ placed on the
axis of symmetry, where
\ben
{2M \over c^2 L}=1.
\een
In this representation,  the event horizon corresponds to the
interval of the axis of symmetry occupied by rod and the mass per
unit length condition guarantees  that the horizon is non-singular
and that there is a smooth extension through it. To obtain Rindler
spacetime, that is flat space in accelerating coordinates, one
takes a semi-infinite rod which is now the acceleration horizon. These
two case are non-singular because there the portion of the axis
not occupied by the rod has no deficit angle. The situation
changes if one considers two disjoint rods. Deficit angles are now
inevitable and one may take them to be along the portions of the
axes running off to infinity. Using the contour integral for the
metric function  $k$ one finds \cite{GibbonsPerry} that the values
are related to the net Newtonian force on one rod due to the
other. Another  case is to take one semi-infinite rod and one
finite rod. This gives the C-metric. Examination of the deficit
angles shows that  to provide the acceleration one must pull the
black hole with a string whose tension cannot exceed the limit.
Note that according to classical General Relativity, which
incorporates Newton's Second Law,  there is no upper bound on the
{\sl acceleration} since we may apply a given force to an
arbitrary small mass. Quantum gravity may impose an upper bound on
acceleration and probably does. This we will turn to briefly later.

\section{Melvin solution}

Another illustration of the principle may be obtained
by considering the Melvin flux tube in Einstein Maxwell theory.
The bundle of flux lines is on the verge of collapsing, being kept up by
Maxwell stresses. One can imagine charges at either end of the flux
tube and one then needs to calculate the force between them.

 The solution is
\ben
ds^2 = -( 1+ {r^2 \over a^2} ) ^2 \Bigl \{ -dt^2 + dz^2 + dr^2 +
{ r^2\over (1+ {r^2 \over a^2 } )^4}  d \phi ^2 \Bigr \}, \label{melvin}
\een
with
\ben
F= B_0 { r dr \wedge d \phi \over (1 + { r^2 \over a^2 } ) ^2 },
\een
and $a= {1  \over B_0 \sqrt{\pi G} }$. For convenience I have set
$c=1$ in ( \ref{melvin}).
The total magnetic flux is $\Phi_m= {G \over B_0}$. The integrated
stress across  the flux tube is proportional to $a^2 B_0^2$
which, restoring units, is proportional to the maximal tension
$ c^4 \over 4G$. The exact factor of proportionality
depends on how one defines  the total force. This is not completely
obvious
in this very non-linear context. It is possible that with an
appropriate
definition even the factor would come out right.

\section{Higher Dimensions}
One may apply the heuristic argument in the introduction in $n$
spacetime dimensions  but one gets a bound on $F/D^{n-4}$. The
method of rods breaks down because there is no analogue of the
Weyl metrics and the C-metric in higher dimensions. Moreover  the
idea of deficit angles does not go over in  a nice way. What are
 natural  sources for distributional Ricci curvature are not strings
(with  two-dimensional world sheets) but rather $(n-2)$-branes
with co-dimension two world volumes. A closely related point is
that one cannot think of the Regge calculus in terms of a network
of strings in higher dimensions. One may see this mis-match by
applying elementary dimensional analysis  to the classical
Einstein equations
\ben
R_{\mu \nu} -{ 1\over 2} R g_{\mu \nu} = { 8 \pi G \over c^4}
T_{\mu \nu}.
\een
The left hand side has the dimensions of $L^{-2}$. The stress
tensor $T_{\mu \nu}$ has the dimensions of stress , i.e $F L
^{(-n-2)}$ or, equivalently,  energy per unit $(n-1)$-volume. Thus
$c^4 \over G$ has dimensions $F L^{-(n-4)}$ and no statement can
be made about fundamental limits on forces or indeed lengths but
just this combination.

\section{String Theory}
It is a striking fact that
in  classical string theory one is given a natural unit of force
or  tension, the energy per unit length of the string \cite{Veneziano}.
\ben
F_s= { 1\over 2 \pi \alpha ^\prime},
\een
where $\alpha ^\prime$ is called the Regge slope parameter.
In fact not everyone defines $\alpha ^\prime$ in this way. Some  prefer
to insert a factors of $\hbar c$ so  that $\sqrt{\alpha^\prime}$
is a fundamental length.
However from the classical  point of view, it
 is more natural to use the tension
$F_s$ rather than $\alpha ^\prime$ as the fundamental constant of
string
theory. We shall see shortly how the  Regge slope parameter
enters the quantum theory. It seems clear, and this is without benefit
of quantum mechanics, that if gravity and string theory are related
then  these two tensions must be proportional, that is purely
classically one must have
\ben
F_g = {1 \over 4} { c^4 \over G} \propto F_s = { 1\over 2 \pi \alpha ^\prime}.
\label{relation}\een

Note that Planck's constant  does not enter (\ref{relation}).
The two theories may be related purely classically.
Related observations have been made by Veneziano \cite{Veneziano}
and are subject to a debate in \cite{trialogue} on the meaning
of ``fundamental constants''. The view taken here is an operational
one. Thus  a possible unit of velocity is that of the universal
upper limit to the speed of propagation of closed string states.
One could use as a unit  the upper limit to the speed of one's favourite
race-horse but that would lack the appealing feature of universality.
It is a non-trivial fact about the world, i.e. a law of nature,
that such a universal limit exists
(at least for closed, as opposed to open string states). Moreover this
law of nature, unlike that of Galilean physics,  is not invariant
under independent re-scalings of any well defined units of
length $L$ and time $T$.

In like
fashion, I claim that it is a non-trivial fact about the world that
there is a natural upper bound for the  tension or force in the
macroscopic world of general relativity  and we could
use that as a natural unit, or fundamental constant.
It then becomes an interesting question of how this is related to
to the natural unit of tension in the classical  micro-world of strings.
The precise relation must involve the string coupling constant
$g_s= e^\Phi$, where $\Phi$ is the classical value of the dilaton.
Since I have nothing new to say about it, factors of $g_s$ will be
ignored in what follows.

At the purely classical level, the invariance of the
equations of  classical general relativity without sources under rescaling
\ben
g_{\mu \nu} \rightarrow \lambda ^2 g_{\mu \nu}, \label{scale}
\een with $\lambda$ a constant, makes it clear that in that theory
there can be   no upper bound
to masses, distances or times and hence no natural or fundamental unit
of mass, length or time, since under (\ref{scale}) all three scale as
$\lambda$
\ben
M \rightarrow \lambda M \qquad L \rightarrow \lambda  L\qquad T
\rightarrow \lambda T \label{dimensions}.
\een The fundamental constants $c$ and $G$ are invariant
as is the   combination $c^4 \over G$ because they
have  the correct dimensions, for example
\ben
\Bigl [{ c^4 \over G} \Bigr ] = ML T^{-2}.
\een
Quantities like angular momentum $J$ and action $S$  satisfy
\ben
\Bigl [J \Bigr ]= \Bigl [S \Bigr ]= ML^2 T^{-1},
\een
and so general relativity provides no complete set of natural units,
 for that we need
Quantum Mechanics and Planck's constant. In fact one may regard
(\ref{scale}) as the residual invariance that is left from
the two parameter groups of scalings of the laws of  special relativity
under which
\ben
M \rightarrow \lambda M\qquad L \rightarrow \mu L\qquad T \rightarrow
\mu T,
\een
and the laws of  Newtonian Gravity under which
\ben
M \rightarrow \lambda M \qquad L \rightarrow \mu L\qquad T \rightarrow
\mu ^{ 3 \over 2} \lambda ^{ -{ 1\over 2}}. \label{kepler}
\een
Note that the non-relativistic scaling (\ref{kepler})
is just the  statement of Kepler's Third Law.

To return to classical string theory:
the Nambu-Goto  action $S$ of a classical string is
given by
\ben
S= F_s \int _\Sigma  dxdt  = { 1\over 2 \pi \alpha ^\prime} \int
_\Sigma  dx dt ,
\een
where $\Sigma $ is the world sheet of the string.

In fact, the contribution to the gravitational action from a conical defect
or cosmic string of the form we contemplated earlier is
identical up to a factor if we make use of (\ref{relation})
\cite{GibbonsPerry}. This confirms the relation (\ref{relation}).
Note finally that the scaling invariance (\ref{dimensions}) of classical general
relativity is  shared by  classical
relativistic  string theory. If it weren't,  then (\ref{relation})
wouldn't make sense.

\section{ Born-Infeld Theory}

One sees the upper bound on the tension arising naturally
in the  Born-Infeld Lagrangian  which is
an  effective theory arising from open strings.
Actually what comes into the Born-Infeld action
is the force on the end of the string because
one adds a boundary term

\ben
\int_{\partial \Sigma}  eA_\mu dx^\mu
\een
where $e$ is called the charge carried by an  end of the string.
However, the effective action involves only the product $e F_{\mu
\nu}$ and is (ignoring a possible additive constant and factors of $g_s$)

\ben
- F^2_s
\sqrt { - {\rm det} \bigl (g_{\mu \nu}  +2 \pi \alpha ^\prime F_{\mu
\nu} \bigl )}
\een

Expanding out to lowest order in $F_{\mu \nu}$
one gets the standard Maxwell Lagrangian but at the full non-linear
level, the Born-Infeld action
 gives rise to an upper bound to the force or tension on the ends
of  the string:

\ben
e E_s= { 1\over 2 \pi \alpha ^\prime} =F_s,
\een
where $ E_s$ is the critical electric field.

\section{Regge Trajectories}

String theory arose in attempt to obtain models with
Regge trajectories. Classically for an open relativistic string, one has
the bound
\ben
J \le { \alpha ^\prime \over c^3}  M^2,
\een
while for a  closed  relativistic  string \cite{Scherk}
\ben
J \le {1 \over 2} { \alpha ^\prime \over c^3} M^2.
\een

 It is interesting to
compare this with the classical cosmic censorship limit on Kerr black holes
\ben
J \le { G \over c} \thinspace M^2.
\een
If, as many people have, one thought of the black hole
as a rotating ring with a tension, one would obtain, up to a factor of
proportionality, the relations (\ref{relation}). This nice
picture fails for higher dimensional rotating black
holes\cite{MyersPerry}
essentially because of the different dependence of the gravitational
force on separation.

\section{Quantum Mechanics}

The introduction by Planck of a
unit of action, or equivalently angular momentum,
breaks the scaling invariance (\ref{scale}) and makes possible
a complete system of fundamental units \cite{Planck}.
 Conventionally, one introduces
 the Planck length,
\ben l_p=
\sqrt{G \hbar \over c^3},\een
which one believes may give a least  length, although
it is probably rather shorter than the string length
\ben
l_s = \sqrt {\alpha ^\prime \hbar c}.
\een

Another way to obtain "fundamental units" pre-dating Planck and
avoiding the
introduction of Planck's constant $\hbar$  is to follow
Stoney, coiner of the term electron, and note that the existence in
nature
 of a
fundamental electric $e$  charge breaks the residual scaling invariance
(\ref{scale})  \cite{Stoney} because classically
\ben \Bigl [ e\Bigl ] = M^{ 1\over 2} L^{ 3 \over 2} T^{ -1} . \een

Of course the in quantum field theory one drops  $G$
but introduces $\hbar$ and to get  a natural unit of charge
$\sqrt{4 \pi \hbar c}$ and obtains the usual dimensionless
gauge coupling constant
$$g_e= { e\over \sqrt{4 \pi \hbar c }}.$$

\section{Maximal Acceleration and Temperature}

In any  special relativistic theory  acceleration $a$ may be
converted to an  inverse length $ac^{-2}$ and in any theory one
may think of temperature $T$ as an energy. In any quantum theory
temperature $T$ and time are related via periodicity in imaginary
time with period $\beta = \hbar T$, In a relativistic quantum
theory one puts these two fact together to get the purely
kinematic Unruh relation
\ben
T= {\hbar a \over 2 \pi c}.\een

It follows that a theory with a minimal length should have a
maximal acceleration and a maximal temperature. By the same token,
a theory with a maximal temperature should have have a maximal
acceleration. There is a clear mechanism here: if one tried to
increase one's  acceleration one would absorb  hotter and hotter
Unruh radiation which would, from the point of view of an inertial
observer, have to be emitted at the expense of the energy source
causing the acceleration.

It is clear that a maximal acceleration or maximal temperature
 can come out of neither classical general relativity
nor classical string theory. It clearly emerges in some form from
classical string theory and quantum gravity. The Hagedorn temperature
and the Hawking temperature of a Planck mass black hole
will, up to factors give its magnitude.  Whether there is a simple
 universal
value however remains unclear. For some ideas about this with some
 references to earlier work the reader is directed to \cite{Schuller}.

\end{document}